\documentstyle[12pt]{article}
\begin{document}
\title{Factorization and superpotential of 
the $PT$ symmetric Hamiltonian}
  \author{V. M. Tkachuk, T. V. Fityo\\
  {\small Ivan Franko Lviv National University, 
         Chair of Theoretical Physics }\\
        {\small 12 Drahomanov Str., Lviv UA--79005, Ukraine}\\
           {\small E-mail: tkachuk@ktf.franko.lviv.ua}}

\maketitle

\begin{abstract}
We study the factorization of the $PT$ symmetric Hamiltonian.
The general expression for the superpotential corresponding to the 
$PT$ symmetric potential is obtained and the explicit examples 
are presented.


\end{abstract}

\section{Introduction}

The $PT$ symmetric complex potentials 
suggested in \cite{Ben98,Ben99J,Ben99R}
have recently attracted a fair amount of attention. 
It was shown that for several  $PT$ symmetric complex 
potentials the spectrum of the corresponding Hamiltonian is real so long as 
the $PT$ symmetry is not spontaneously broken. 
Just this feature is the main  
reason of interest to them.
Recently it has been proved that supersymmetric (SUSY) quantum  mechanics 
(for a review see \cite{Coo95})
is a useful tool for the 
investigation of the eigenvalue problem not only for the 
Hermitian Hamiltonian 
but also for the non-Hermitian Hamiltonian with a complex potential 
\cite{Can98,And99,Zn00,Bag269,BagL1,Can01,Zn01p,Bag01,Dor01}.

In this paper we shall answer the following question. 
What general expression 
for the superpotential leads to the $PT$ symmetric potentials. 
This gives us a 
possibility to obtain a general expression for the quasi-exactly 
solvable (QES) $PT$ 
symmetric potential for which we know in explicit form one eigenstate. 
In this connection
it is worth to note that the nature of $PT$ symmetric QES potentials
of a special form has been studied in \cite{Ben273,Kha272,Bag095} 
(see also references therein).

Note that there is no problem to generate the 
QES arbitrary complex potential with one known eigenstate. 
Even for the case of two or three  eigenstates it is 
possible to obtain a general expression for 
a complex QES potential using the 
supersymmetric method proposed in \cite{Tka98,Tka99Ph}
when the $PT$ symmetry is not imposed.
But when the $PT$ symmetry is imposed on the potential then it 
is not a trivial 
problem to a obtain a general expression for the QES potential even with 
one known eigenstate. 
Only this problem is considered in the paper.

\section{Superpotential of the $PT$ symmetric potentials}

Let the Hamiltonian read as
\begin{equation}\label{Ham1}
H=-\frac{1}{2}\frac{d^2}{dx^2}+V(x),
\end{equation}
where $V(x)=V_1(x)+iV_2(x)$ is a complex potential.
The Hamiltonian is called $PT$ symmetric when
\begin{equation}
PTH=HPT,
\end{equation}
where $P$ is the parity operator acting as the spatial 
reflection: $Pf(x)=f(-x)$, 
and $T$ is the complex conjugation operator: $Tf(x)=f^*(x)$. 
In the explicit form 
the condition of the $PT$ symmetry for a potential $V(x)$ reads
\begin{equation}\label{PTPot}
V^*(-x)=V(x).
\end{equation}

Suppose that the Hamiltonian can be written in the factorized form
\begin{equation}\label{Hfact}
H=\frac{1}{2}\left(-\frac{d}{dx}+W\right)\left(\frac{d}{dx}+W
\right)+\varepsilon=
-\frac{1}{2}\frac{d^2}{dx^2}+\frac{1}{2}(W^2-W')+\varepsilon,
\end{equation}
where $W$ is the so called superpotential,
$W'=dW/dx$, $\varepsilon$ is the energy of 
factorization. Note that in our case $W$ and $\varepsilon$ are 
complex valued. 
The wave function corresponding to the energy $\varepsilon$ reads
\begin{equation}\label{psi}
\psi_{\varepsilon}=Ce^{-\int Wdx}.
\end{equation}

Using (\ref{Ham1}) and (\ref{Hfact}) we obtain the relation between 
the potential and the superpotential
\begin{equation}\label{potf}
V(x)=\frac{1}{2}(W^2-W')+\varepsilon.
\end{equation}
Then the $PT$ symmetry condition (\ref{PTPot}) reads
\begin{equation}\label{cond1}
(W^{*}(-x))^2+\frac{d}{dx}W^*(-x)+2\varepsilon^*=W^2(x)-\frac{d}{dx}W(x)
+2\varepsilon.
\end{equation}

Thus the potential $V(x)$ is $PT$ symmetric when the 
superpotential satisfies condition (\ref{cond1}). 
To solve this equation we rewrite it in the 
following form
\begin{equation}\label{neweq}
U'_+(x)=U_+(x)U_-(x)+2(\varepsilon-\varepsilon^*),
\end{equation}
where
\begin{eqnarray}\label{uplus}
U_+(x)=W(x)+W^*(-x), \\ \label{uminus}
U_-(x)=W(x)-W^*(-x).
\end{eqnarray}
As follows from (\ref{uplus}) and (\ref{uminus}) $U_+$ is $PT$ symmetric and 
$U_-$ is anti $PT$ symmetric
\begin{eqnarray}\label{ptuplus}
U_+^*(-x)=U_+(x), \\
U_-^*(-x)=-U_-(x).
\end{eqnarray}

Equation (\ref{neweq}) can be easily solved with respect to $U_-$ 
for a given $U_+$ or vice versa. 
We use the solution with respect to $U_-$, i.e.,
\begin{equation}
U_-=\frac{{U'}_+-2(\varepsilon-\varepsilon^*)}{U_+}.
\end{equation}
Then from (\ref{uplus}) and (\ref{uminus}) we obtain
\begin{equation}\label{spot}
W=\frac{1}{2}\left\{
U_++\frac{{U'}_+-2(\varepsilon-\varepsilon^*)}{U_+}
\right\}.
\end{equation}
This expression for the superpotential 
is the main result of this paper.
It is interesting to note that equation (\ref{neweq})
and the expression for superpotential (\ref{spot}) 
are formally similar to the equation and superpotential 
obtained in our papers \cite{Tka98,Tka99Ph} 
where we studied the real QES potential
with two known eigenstates.

Substituting (\ref{spot}) into (\ref{potf}) we obtain the $PT$ symmetric 
potential which can be written in the following form
\begin{equation} \label{Vupm}
V(x)={1\over 8}(U_+^2+U_-^2)-{1\over 4}U'_- 
+{1\over 2}(\varepsilon^* +\varepsilon).
\end{equation} 
Note that for this potential we know in the explicit form at least 
one level $\varepsilon$ and the corresponding wave function (\ref{psi}). 
This function corresponds to the discrete spectrum when 
it vanishes at infinity
or to the continuum spectrum when it is restricted.
In these cases potential (\ref{Vupm}) can be called 
the $PT$ symmetric QES potential with one known eigenstate.
It is also possible that (\ref{psi}) will not satisfy the necessary
conditions. Then this function does not belong to the eigenfunctions
of the Hamiltonian.

Now let us consider a case of the $PT$ symmetric wave function, 
namely, $PT\psi_0=\psi_0$. In this case we have 
\begin{equation} \label{PTW}
W^*(-x)=-W(x)
\end{equation}
and thus $U_+=0$. 
Therefore, in order to use equation (\ref{spot}) for this case 
we put $U_+=\alpha f(x)$, where $\alpha$ tends to zero. 
Then from (\ref{spot}) we obtain
\begin{equation}
W(x)=\frac{1}{2}\left\{\frac{f'}{f}-B\frac{i}{f}\right\},
\end{equation}
where 
$$B=\lim_{\alpha\to 0}2\frac{\varepsilon-\varepsilon^*}{\alpha}.$$
We see that the imaginary part of energy must tend to zero. 
Thus, we show that (\ref{spot}) in the special case (\ref{PTW})
reproduces the well known result. 
Namely when the wave function is $PT$ symmetric
the eigenvalue is real.

\section{Examples}

To illustrate  the described method  we give two explicit examples 
of the $PT$ symmetric potentials. 
All expressions depend on the function $U_+(x)$, which
can be called a generating function. We may choose various functions $U_+(x)$
and obtain as a result various $PT$ symmetric potentials. 
In the considered 
examples we specially choose such a function $U_+(x)$ which leads  
to the proper eigenfuction. 
Therefore, the obtained $PT$ symmetric potentials are the QES ones with
one known eigenstate.

{\em Example 1}

Let us consider
\begin{equation}
U_+=\frac{i\alpha }{(x+ia)^n},
\end{equation}
where $n$ is an odd number. Then
\begin{equation}
U_-=-\frac{n}{x+ia}-\frac{4\mbox{Im}\varepsilon}{\alpha}(x+ia)^n,
\end{equation}
\begin{equation}
W=\frac{i\alpha }{2(x+ia)^n}-\frac{n}{2(x+ia)}-\frac{2\mbox{Im}\varepsilon}{\alpha}
(x+ia)^n,
\end{equation}
\begin{equation}
V=\mbox{Re}\varepsilon-\frac{\alpha^2}{8}\frac{1}{(x+ia)^{2n}}+
2\frac{(\mbox{Im}\varepsilon)^2}{\alpha^2}(x+ia)^{2n}+\frac{1}{8}
\frac{n^2-2n}{(x+ia)^2}+2\frac{n\mbox{Im}\varepsilon}{\alpha}(x+ia)^{n-1}
\end{equation}
If $n>1$ then we can write the wave function in the following form
\begin{equation}
\psi_{\varepsilon}=C(x+ia)^{n/2}\exp\left(
\frac{i\alpha}{2(n-1)}\frac{1}{(x+ia)^{n-1}}+
2\frac{\mbox{Im}\varepsilon}{\alpha}\frac{(x+ia)^{n+1}}{n+1}
\right)
\end{equation}
or if $n=1$, it reads
\begin{equation}
\psi_{\varepsilon}=C(x+ia)^{(1-i\alpha )/2}\exp\left(
\frac{\mbox{Im}\varepsilon}{\alpha}(x+ia)^{2}
\right).
\end{equation}
To obtain a bound state we must set $\frac{\mbox{Im}\varepsilon}{\alpha}<0$.

Note that in the case of $n=1$ we have the $PT$ symmetric harmonic oscillator 
with a regularized centrifugal-like core which is exactly solvable 
\cite{Zn220, L7165}.
Thus, the considered example generalizes the 
$PT$ symmetric harmonic oscillator to the
quasi-exactly solvable $PT$ symmetric potential with one known eigenstate.

It is interesting to stress that in the limit  $\mbox{Im}\varepsilon\to 0$, 
$\alpha\to 0$ and $\lim2\frac{\mbox{Im}\varepsilon}{\alpha}=B=const$ we obtain 
the $PT$ symmetric wave function with a real eigenvalue. 
This just confirms the 
result obtained in the end of the previous section.

{\em Example 2}

This example represents the periodic $PT$ symmetric potential. Let us take 
\begin{equation}
U_+=\alpha e^{ikx}.
\end{equation}
Then we obtain
\begin{equation}
U_-=ik-4i\frac{\mbox{Im}\varepsilon}{\alpha}e^{-ikx},
\end{equation}
and the superpotential, potential and wave function read, respectively,
\begin{equation}
W=\frac{\alpha}{2}e^{ikx}+\frac{ik}{2}-2i\frac{\mbox{Im}\varepsilon}{\alpha}
e^{-ikx},
\end{equation}

\begin{equation}
V=\mbox{Re}\varepsilon-\frac{k^2}{8} 
+\frac{\alpha^2}{8}e^{2ikx}-2\frac{(\mbox{Im}\varepsilon)^2}{\alpha^2}e^{-2ikx}+
2k\frac{\mbox{Im}\varepsilon}{\alpha}e^{-ikx},
\end{equation}

\begin{equation}
\psi_{\varepsilon}=\exp\left(
-\frac{ikx}{2}+\frac{i\alpha}{2k}e^{ikx}-2\frac{\mbox{Im}\varepsilon}
{\alpha k}e^{-ikx}\right).
\end{equation}

In the case of $\mbox{Im}\varepsilon=0$ this QES potential becomes exactly solvable
and corresponds to the potential studied in \cite{Can98}.

\section{Conclusions}

We have obtained the general expression for superpotential (\ref{spot})
which corresponds to the $PT$ symmetric potential (\ref{Vupm}).
The $PT$ symmetric function $U_+(x)$ plays the role of the generating
function. Choosing different functions $U_+(x)$ we obtain different 
superpotentials (\ref{spot}) and the corresponding $PT$ symmetric
potentials (\ref{potf}) or (\ref{Vupm}).
This $PT$ symmetric potentials can be called as QES potentials
with one known eigenfunction (\ref{psi}) and the corresponding energy 
$\varepsilon$.
Of course, the solution (\ref{psi}) must satisfy necessary conditions
in order to be the eigenfunction of the Hamiltonian. 

In the considered examples we specially choose the $U_+(x)$
which lead to proper eigenfunctions. 
The potentials considered in examples 1 and 2
are interesting because at some values of the parameters 
they become exactly solvable potentials which were 
studied earlier. Thus, the considered  examples 
generalize the exactly solvable $PT$ symmetric potentials to the QES  
potentials with one known eigenstate.

\end{document}